\documentclass[fleqn,usenatbib]{mnras}

\usepackage{newtxtext,newtxmath}
\usepackage[T1]{fontenc}
\usepackage{ae,aecompl}

\usepackage{hyperref}

\def\xlinkspace#1 #2{%
\ifx\relax#2%
\xlinkdash#1-\relax
\else
\xlinkdash#1 -\relax
\expandafter\xlinkspace\expandafter#2%
\fi}

\def\xlinkdash#1-#2{%
\ifx\relax#2%
\tmp{#1}%
\else
\tmp{#1-}%
\expandafter\xlinkdash\expandafter#2%
\fi}

\usepackage{graphicx}	% Including figure files
\usepackage{amsmath}	% Advanced maths commands
\usepackage{graphics,graphicx}
\usepackage{epstopdf}
\usepackage{color}
\usepackage{deluxetable}
\usepackage{natbib}
\def\apj{\rm ApJ}
\def\apjl{\rm ApJL}
\def\apjs{\rm ApJS}
\def\aj{\rm AJ}
\def\mnras{\rm MNRAS}
\def\nat{\rm Nature}

\def\aap{\rm A\&A}
\newcommand{\msun}{M$_{\odot}$}

\title[Off-center Black Holes in Dwarf Galaxies]{The Origins of Off-Centre Massive Black Holes in Dwarf Galaxies}

\author[Bellovary et al.]{Jillian M. Bellovary$^{1,2,3}$\thanks{E-mail: jbellovary@amnh.org}, Sarra Hayoune$^4$,  Katheryn Chafla$^5$,  Donovan Vincent$^6$,\newauthor Alyson Brooks$^{7}$,   Charlotte R. Christensen$^{8}$,  Ferah D. Munshi$^{9}$,  Michael Tremmel$^{10}$,  \newauthor Thomas R. Quinn$^{11}$, Jordan Van Nest$^{9}$, Serena K. Sligh$^{9}$, Michelle Luzuriaga$^{12}$  \\
$^{1}$Department of Physics, Queensborough Community College, City University of New York, 222-05 56th Ave, Bayside, NY, 11364\\
$^{2}$Department of Astrophysics, American Museum of Natural History, Central Park West at 79th Street, New York, NY 10024, USA\\
$^{3}$Department of Physics, Graduate Center, City University of New York, New York, NY 10016, USA\\
$^{4}$Department of Physics, Stevens Institute of Technology, Hoboken, NJ, 07030, USA\\
$^{5}$University of Connecticut, Storrs, CT, 06269, USA\\
$^{6}$Manhattan College, Riverdale, NY, 10471, USA\\
$^{7}$Department of Physics \& Astronomy, Rutgers University, 136 Frelinghuysen Rd., Piscataway, NJ 08854, USA\\
$^{8}$Department of Physics, Grinnell College, 1115 8th Ave, Grinnell, IA, 50112, USA\\
$^{9}$Department of Physics and Astronomy, University of Oklahoma, 440 W. Brooks St, Norman, OK, 73019, USA\\
$^{10}$Department of Astronomy, Yale University, 52 Hillhouse Ave, New Haven, CT, 06511, USA\\
$^{11}$Department of Astronomy, University of Washington, Box 351580, Seattle, WA 98122, USA\\
$^{12}$]Queens College, City University of New York, Queens, NY, 11367
}
\date{Accepted XXX. Received YYY; in original form ZZZ}

\pubyear{2021}

\begin{document}
\label{firstpage}
\pagerange{\pageref{firstpage}--\pageref{lastpage}}
\maketitle

\begin{abstract}
Massive black holes often exist within dwarf galaxies, and both simulations and observations have shown that a substantial fraction of these may be off-center with respect to their hosts.  We trace the evolution of off-center massive black holes (MBHs) in dwarf galaxies using cosmological hydrodynamical simulations, and show that the reason for off-center locations is mainly due to galaxy-galaxy mergers.  We calculate dynamical timescales and show that off-center MBHs are unlikely to sink to their galaxys' centers within a Hubble time, due to the shape of the hosts' potential wells and low stellar densities.  These wandering MBHs are unlikely to be detected electromagnetically, nor is there a measurable dynamical effect on the galaxy's stellar population.  We conclude that off-center MBHs may be common in dwarfs, especially if the mass of the MBH is small or the stellar mass of the host galaxy is large.  However detecting them is extremely challenging, because their accretion luminosities are very low and they do not measurably alter the dynamics of their host galaxies.

\end{abstract}

\begin{keywords}
black hole physics  -- galaxies: dwarf
\end{keywords}

%%%%%%%%%%%%%%%%%%%%%%%%%%%%%%%%%%%%%%%%%%%%%%%%%%

%%%%%%%%%%%%%%%%% BODY OF PAPER %%%%%%%%%%%%%%%%%%

%%% papers to look at and possibly cite before re-submit:
% uddipan banik, dynamical friction and core stalling:  https://arxiv.org/abs/2103.05004
% same author, sent to me by same: https://academic.oup.com/mnras/article/483/2/1558/5222688?login=true
% (but the above is about lensing)
% Pierre Boldrini sent two:  https://arxiv.org/abs/2003.02611  and   https://arxiv.org/abs/2007.03010  VERY GOOD

\section{Introduction}

Observational evidence has been mounting for the existence of massive black holes (MBHs) in dwarf galaxies, which inform our understanding of MBH-host galaxy scaling relations as well as provide clues to the origins of supermassive black hole (SMBH) seeds.  These MBH candidates have been discovered in several ways including via X-rays \citep{Pardo16,Baldassare17,Mezcua18,Birchall20}, broad and narrow emission lines \citep{Reines13,Moran14,Chilingarian18,Dickey19,Cann20}, infrared emission   \citep{Satyapal14}, nuclear variability \citep{Baldassare18, Lira20}, radio emission \citep{Mezcua19,Reines20}, and masers \citep{Zaw20}.  As the number of candidates increases, one can begin to compile a sample and determine whether overall these these lower--mass counterparts of massive galaxies fall on the same scaling relations, such as $M_{\rm BH}-\sigma$, $M_{\rm BH}-M_{\rm bulge}$, and $M_{\rm BH}-M_{\rm star}$.  While some efforts portend that the $M_{\rm BH}-\sigma$ relation does not change as masses decrease \citep{Baldassare20}, others find evidence of a low-mass downturn of the $M_{\rm BH}-M_{\rm bulge}$ relation \citep{Graham15,Mezcua17}.  See \citet{Greene20} for a thorough review on this topic.

MBHs in dwarf galaxies are not always straightforward to detect, however.  In our prior work, \citet{Bellovary19} found that approximately half of MBHs in simulated dwarf galaxies are not located in the nuclear regions of their galaxies.  This result was confirmed observationally by \citet{Reines20}, who executed a radio survey of 111 dwarf galaxies detected in FIRST.  Of their sample of dwarf galaxies with compact radio sources consistent with AGN activity, approximately half are off-center.   \citet{Mezcua20} also show evidence for off-center AGN activity in dwarfs, as indicated by spatially resolved emission lines.

Off-center MBHs will have different dynamical and accretion histories than their nuclear counterparts, and will thus have different effects on their host galaxies.  Since these MBHs spend a large fraction of their orbits in less dense regions, they will undergo fewer accretion events.  Thus, host galaxies may experience less quenching of star formation, due to the lack of radiation feedback due to accretion onto the black hole.  The MBHs themselves will be more difficult to detect electromagnetically, since their accretion luminosities will be low.  Gravitational waves from MBH-MBH mergers in dwarf galaxies will also be rare if one or more MBH is off-center; the odds of the black holes forming a bound pair are decreased.

These observational challenges have important repercussions.  The difficulty of detecting these MBHs will result in an underestimate of the occupation fraction of MBHs in dwarf galaxies as well as the overall density of MBHs in the universe.  Since dwarf galaxies are the most numerous type of galaxy, this underestimate may be substantial.  Statistical measurements of occupation fractions (as well as more direct measurements of black hole masses) are important because they can aid in constraining the formation mechanism(s) of SMBH seeds \citep{Greene12,VolonteriNatarajan,Miller15}, as can gravitational wave measurements by future observatories such as the Laser Interferometer Space Antenna (LISA) \citep{Ricarte18}.  Most theoretical predictions of these quantities assume central MBHs, and contain corresponding assumptions about their detectability.  It is thus imperative that we understand and quantify the off-center fraction.

In this paper we explore the origins and detectability of off-center MBHs in dwarf galaxies.  In Section \ref{sect:sims} we describe the simulations used in our study.  We examine how MBHs become off-center in Section \ref{sect:how}, and why they remain off-center in Section \ref{sect:why}.  In Section \ref{sect:detect} we will discuss whether off-center MBHs can be detected, and we summarize our results in Section \ref{sect:summary}.

\section{Simulations}\label{sect:sims}

We use the state-of-the-art N-Body Tree + Smoothed Particle Hydrodynamics (SPH) code ChaNGa \citep{Menon15} to simulate realistic dwarf galaxies.  This code as well is its predecessor GASOLINE \citep{Stadel01, Wadsley04} uses modern methods and realistic sub-grid models to reproduce a range of observed galaxy properties using high-resolution simulations \citep{Anderson17, Munshi17,Tremmel18,Bellovary19,Sanchez19,Wright20,Applebaum20,Akins20}.  The particular simulations in this work are  known as the MARVEL-ous Dwarfs and the DC Justice League, and are described in detail in \citet{Bellovary19}; we summarize them here.   

\subsection{Simulation Properties}

All of our simulations use a modern SPH kernel calculation, which uses a geometric mean density in the SPH force expression and accurately reproduces shearing flows \citep{Ritchie01, Menon15, Wadsley17}.  They include a uniform UV background \citep{Haardt12} and metal diffusion and cooling \citep{Shen10}.  Star formation is determined by the molecular gas fraction as described in \citet{Christensen12}.  Stars form probabilistically when the gas density surpasses a threshold of  $\rho_{\rm th} > 0.1$ cm$^{-3}$, although the actual density of star-forming gas is usually 10-1000 cm$^{-3}$ due to the molecular gas requirement.  In addition the gas temperature must be $T < 1000$ K.  We set the star formation efficiency parameter equal to $c_* = 0.1$, and stars form with a Kroupa IMF \citep{Kroupa}.  Supernovae release $E_{SN} = 1.5 \times 10^{51}$ erg in accordance with the blastwave model described in \citet{Stinson06}, which disables cooling for the theoretical lifetime of the momentum conserving phase.

Our simulations were selected from two different uniform volumes, using the ``zoom-in'' technique \citep{Katz93}.   The suite known as the ``MARVEL-ous Dwarfs'' (Captain Marvel, Elektra, Rogue, and Storm) \citep{Munshi21}  was selected from a 25 Mpc volume, has a force softening resolution of 60 pc, and particle masses of $M_{\rm dark} = 6660$\msun, $M_{\rm gas} = 1410$\msun, and $M_{\rm star} = 442$\msun~ (``mint'' resolution as described in \citet{Applebaum20}).  It uses a WMAP 3 cosmology \citep{WMAP3} and consists of 65 galaxies.  The other suite, known as the ``DC Justice League'' (Sandra, Ruth, Sonia and Elena), was selected from a 50 Mpc volume,  has a force softening resolution of 170 pc, and particle masses of $M_{\rm dark} = 4.2 \times 10^4$\msun, $M_{\rm gas} = 2.7 \times 10^4$\msun, and $M_{\rm star} = 8000$\msun~ (``near-mint'' resolution).  It uses a Planck cosmology \citep{Planck13} and consists of 113 dwarf galaxies.  Dwarf galaxies are defined as having resolved star formation histories (star formation spanning more than 100 Myr), and minimum and maximum total masses of $10^7$ and $2.5 \times 10^{11}$\msun, respectively.  We identify galaxies using the Amiga Halo Finder \citep{Gill04,Knollmann09} which identifies halos as spherical  spherical regions within which the density satisfies the redshift-dependent overdensity criterion approximated by \citet{Bryan98}.

The two sets of simulations (MARVEL-ous Dwarfs and DC Justice League) differ in both their resolutions and the cosmologies used. Cosmology will have a minor effect compared to environmental factors on properties such as MBH  and star formation, and we do not anticipate any issues due to the difference.
 All recipes for star formation, feedback, and black hole physics (including the seed mass) are the same, regardless of resolution.    There are no systematic differences between the two datasets;  dwarfs in both suites lie along the same observed scaling relations such as stellar mass - halo mass, size-luminosity, and mass-metallicity \citep{Munshi21}.    
  
\subsection{Black Hole Physics}\label{sect:BHphysics}

All black hole physics in our simulations is described in detail in \citep{Bellovary19}; here we present a summary.  Black hole particles form from collapsing, cool ($T < 2 \times 10^4$ K),  low-metallicity ($Z < 10^{-4}$) gas which has a low molecular gas content ($f_{H_2} < 10^{-4}$).  These criteria are meant to broadly represent the direct collapse scenario \citep[e.g.][]{Oh02,Begelman06,Lodato06}.  The threshold density for black hole formation is 3000 cm$^{-3}$ for the lower resolution runs, and $1.5 \times 10^4$ cm$^{-3}$ in the higher resolution simulations.   The gas particle  must  also  exceed  a  Jeans  Mass  criterion, which ensures that the particle is in a region which is likely to collapse.  Once formed, the MBH particle accretes mass from the surrounding gas, until it either depletes its neighbourhood (i.e. one softening length) of gas or reaches a mass of 50,000 \msun, whichever happens first.  In some cases in the Justice League simulations, the neighborhood is depleted first, resulting in black holes with masses of $\sim 25,000$ \msun.

We do not resolve the scales at which MBH-MBH pairs coalesce and merge.  When our particles form a close pair, which involves them having a separation far below our resolution limit, we treat them as a single particle.  Specifically, they must have an interparticle distance of less than two softening lengths, and also meet the criterion $\frac{1}{2}\Delta {\vec{ \rm v}} < \Delta {\vec{ \rm a}} \cdot \Delta {\vec{ \rm r}}$, where $\Delta {\vec{ \rm v}},  \Delta {\vec{ \rm a}}$ and  $\Delta {\vec{ \rm r}}$ represent the relative velocity, acceleration, and radius vectors of the two MBHs respectively.  This criterion mimics the unresolved condition of the two MBHs being gravitationally bound to each other.   The actual coalescence timescale is dependent on environmental factors and can vary greatly, but is expected to be around $10^7 - 10^8$ years \citep{Armitage02, Haiman09, Colpi14, Holley-Bockelmann15}.  These timescales are small compared to the timescales of the simulation; while this caveat must be kept in mind, the broader results of merging MBHs still hold.

Black holes can also grow via the accretion of gas.  We use a modified Bondi-Hoyle prescription which takes into account the angular momentum of the surrounding gas, and prioritizes the accretion of gas with low angular momentum.  If gas is rotationally dominated, the accretion rate is effectively reduced.  See \citet{Tremmel17}  for a more detailed description of the model.  Our accretion rate $\dot M$ is calculated by the following equation:
 
 \begin{equation}\label{eqn:mdot}
 \dot M =\left\{
  \begin{array}{@{}ll@{}}
 \frac{\pi G^2 \alpha M_{\rm BH}^2 \rho}{(v_{\rm bulk}^2 + c_s^2)^{3/2}}, & \text{for $v_{\rm bulk} > v_{\theta}$} \\
 \frac{\pi G^2 \alpha M_{\rm BH}^2 \rho c_s}{(v_{\theta}^2 + c_s^2)^2}, & \text{for $v_{\rm bulk} < v_{\theta}$} \\
   \end{array}\right.
 \end{equation}
 
 \noindent
 where $M_{\rm BH}$ is the black hole mass, $\rho$ is the local gas density, $v_{\rm bulk}$ is the bulk velocity of the gas, $v_{\theta}$ is the tangential velocity of the gas, $c_s$ is the local sound speed, and $\alpha =1 $ if the local gas density $\rho$ is less than the star formation threshold density $\rho_{\rm th}$, and $\alpha = (\rho/\rho_{\rm th})^2 $ if the gas density is greater. 

Thermal feedback due to accretion is injected to the nearest 32 particles along the SPH kernel. We disable cooling for the heated particles over the duration of each particle's own timestep (typically $10^3-10^4$ yr) which mimics the continuous deposition of feedback energy.  The feedback energy is proportional to the mass accretion rate $\dot M$, assuming a radiative efficiency of $\epsilon_r = 0.1$ and a feedback coupling efficiency of $\epsilon_f$ = 0.02.  All accretion and feedback parameters have been rigorously tested \citep[see][]{Tremmel17} and produce galaxies which match observed scaling relations such as $M_{\rm BH} - \sigma$, stellar mass - halo mass, and mass-metallicity.  We also point out that the MBHs in the dwarf galaxies we present here undergo very little accretion, due to their wandering nature, so effects due to our choice of subgrid models will be minimal.

A critical component of our black hole model is dynamical friction (DF).  This force acts to ``drag'' black holes, which are far more massive than the sea of stars they exist within, to galaxy centers (and keep them there).  Cosmological simulations do not resolve the small scales at which DF acts, because the particle sizes are fairly comparable (within an order of magnitude of each other).  We have implemented a subgrid DF model \citep{Tremmel15} based on the Chandrasekhar formula \citep{Chandrasekhar43, Binney08} which estimates the density of stars and dark matter within the unresolved space and adds an additional acceleration, acting as drag on the MBH.  This mechanism allows us to accurately trace the dynamics of MBHs in galaxies, which become complicated during mergers and interactions.  Black holes which become off-center experience realistic orbital timescales, which may or may not be shorter than a Hubble time.  We do not include gas particles in the dynamical friction calculation; due to the collisional nature of gas, it may exhibit different dynamical behaviours compared to collisionless particles.  The most realistic portrayal of the galactic potential is reflected by the particles not subject to hydrodynamical forces.

 Dynamical friction sinking timescales are dependent on black hole mass, and we acknowledge that the seed masses of IMBHs in the early universe are unknown to orders of magnitude.  In these simulations, we form black holes via rapid collapse and accretion of gas, which occurs in the most overdense regions.  At times multiple BH particles form at once in the same region, which results in their immediate merger.  This process creates an approximate initial mass function (see Figure 1 of \citet{Bellovary19}) with initial black hole masses ranging from the simulation seed mass ($2.5-5 \times 10^4$\msun)  to above $10^6$\msun.  These values bracket the high end of the range of theorized seed masses.  High masses result in more efficient dynamical friction, so an overestimate of seed masses would {\em overestimate} the number of central MBHs.  The fraction of wandering MBHs in dwarfs presented here can thus be seen as a lower limit to the true number, which could be much higher if the black hole seeds were orders of magnitude less massive.
 
 We also note that there are a handful of black holes in our simulations which do not meet the minimum mass criterion for resolved DF, which is $M_{\rm BH} > 3 M_{\rm dark}$.  We exclude these objects from all analysis.

\section{Departure from Galaxy Centers}\label{sect:how}

\begin{figure}
\includegraphics[width=.45\textwidth]{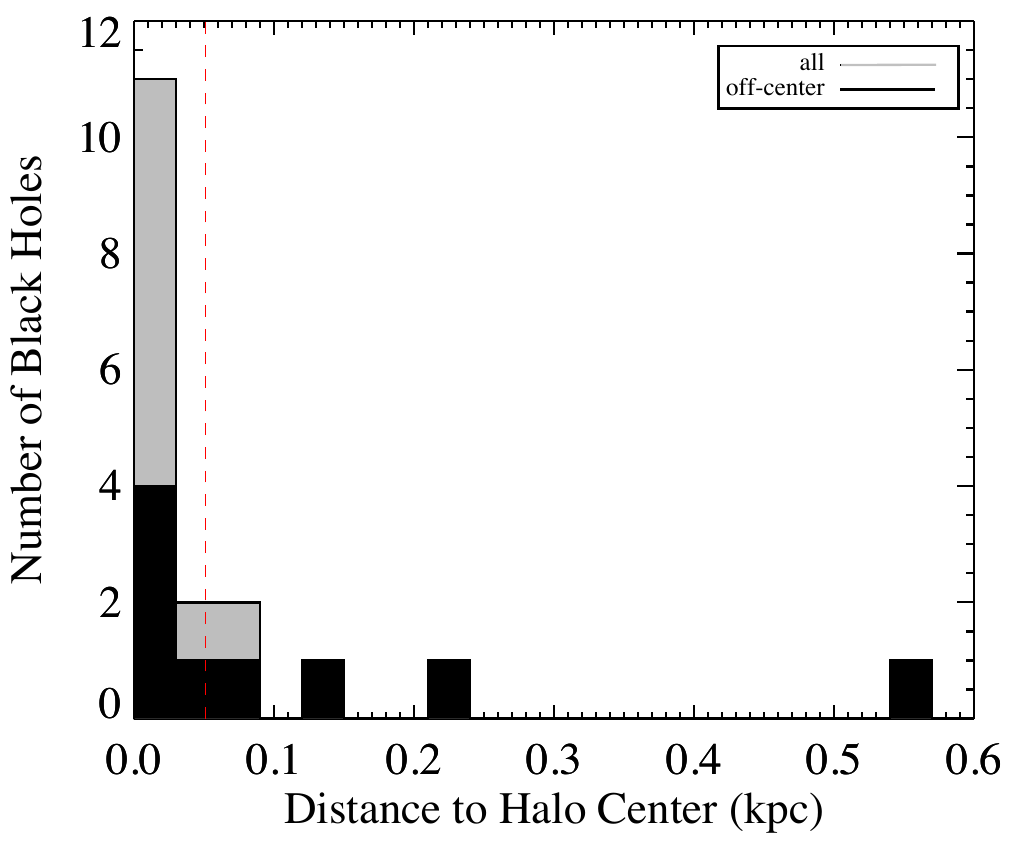}
\caption{Distribution of the physical distance at which MBH seeds form from their halo centers, within 50 Myr of their birth.  The grey histogram  represents all MBHs which are in dwarfs at $z = 0$, and the black histogram line represents those which are off-center at $z = 0$.  The vertical red dashed line indicates $3 \times$ the spatial resolution of the Justice League simulations.  Most MBHs form within the resolution limit, however there are a few which form outside.  Overall the distribution of MBHs which are central and off-center at $z = 0$ have the same distribution at formation.  Only one black hole forms far from its hosts' center (Rogue 1).
\label{fig:centers}}
\end{figure}

MBHs form in small halos ($10^8 < M_{\rm halo} < 10^9$\msun, \citet{Bellovary19}) at very early times ($z < 6$).  The majority of MBHs form in their galaxy centers.  We emphasize that the black holes do not ``know'' where the galaxy center is, and are not placed (or fixed) there artificially.  They simply form where the gas is densest and collapsing, and meets the criteria for seed formation.  In Figure \ref{fig:centers} we show the distribution of distances to galaxy centers shortly after seed formation.  Galaxy centers are found using a shrinking-sphere methodology.  The time after seed formation varies, because simulation snapshots are saved every $\sim 50$ Myr, while seed MBHs can form on very small timesteps; the distance shown here is recorded less than $\sim 50$ Myr after the MBH forms.  The MBHs which eventually become off-center are depicted with a black histogram, while all MBHs in dwarfs are depicted with the grey histogram.  The red dashed line represents our spatial resolution.  Most MBHs form within this range, though a few form outside.  Overall, the distributions are equivalent the two MBH populations, with the exception of Rogue 1, which is discussed below.

\begin{figure*}
\includegraphics[width=.9\textwidth]{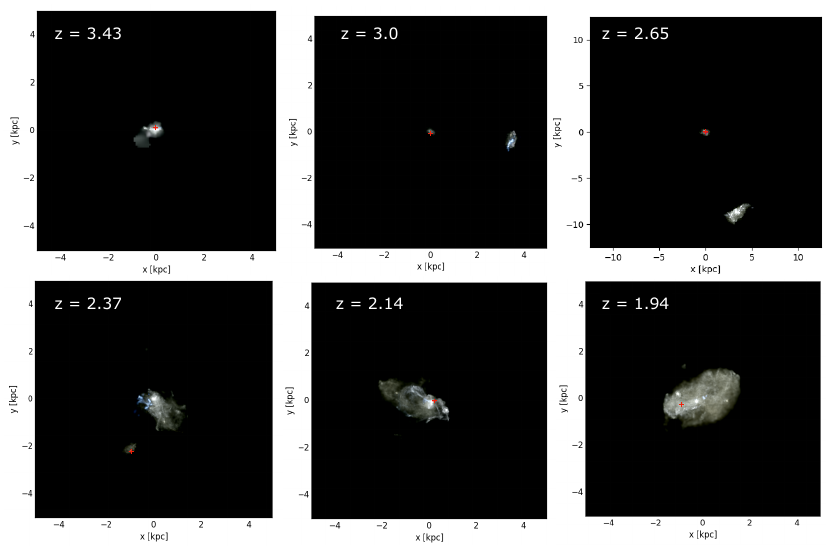}
\caption{Simulated SDSS $rgb$ images of a galaxy merger in which a very small galaxy hosting an MBH (designated by a red cross) merges with a larger galaxy, resulting in an off-center MBH in the larger galaxy.  Redshifts are noted in the top left panel of each images.
\label{fig:pictures1}}
\end{figure*}

The host galaxies subsequently grow by smooth accretion and/or mergers with other galaxies.  As we will demonstrate in this paper, MBHs tend to remain central for galaxies which grow primarily by smooth accretion.  Off-center black holes are likely to be found in galaxies which experience mergers; the most common scenario consists of a minor merger in which the {\em smaller} galaxy hosts the MBH.  As this galaxy collides with a larger galaxy, the MBH's host is tidally stripped, leaving it to wander in the remnant's halo. We define off-center MBHs as existing more than 400 pc from the galaxy center at $z = 0$, which is over twice our force resolution in the lower-resolution simulations.  An example of this scenario is depicted in Figure \ref{fig:pictures1} , where we show simulated images of a galaxy in the Storm simulation.   At  $z = 3.4$, the tiny dwarf is isolated and hosts a central MBH.  In the next two snapshots, one can see a more massive neighbor galaxy approaching.  At $z = 2.14$ the galaxies merge, and by $z = 1.94$ one can see the merger is complete, with the MBH located toward the left edge of the galaxy.    This scenario of the smaller galaxy hosting the MBH is common to all merger-caused offcenter MBHs.  There is one case where the larger host {\em also} hosts a MBH; the galaxy Storm 2 (seen in Figure  \ref{fig:pictures1}) accretes another just after 10 Gyr.  Since the existing MBH is already wandering, there is no noticeable effect when the additional MBH is added.  

%In Figure \ref{fig:pictures2},  a different tiny dwarf hosting another MBH is shown spiraling into the more massive galaxy (which already hosts the aforementioned MBH, which is now marked by a  blue cross).    The galaxies begin to converge at $z = 0.04$, and by $0.02$ (final panel) we see the final state of this galaxy.  This dwarf ends up with two off-center MBHs at $z = 0$, one which is a very recent acquisition.

%%%%  pictures of Storm BH merger 1  %%%%%%%%%%%

%%%%%  pictures of Storm BH merger 2  %%%%%%%%%%%
%\begin{figure*}
%\includegraphics[width=.9\textwidth]{plots/Storm_secondmerger.pdf}
%\caption{Simulated SDSS $rgb$ images of a galaxy merger in which a very small galaxy hosting an MBH (designated by a red cross) merges with a larger galaxy, already hosting an MBH (denoted by a blue cross, see Figure \ref{fig:pictures1}), resulting in an additional off-center MBH in the larger galaxy.  Redshifts are noted in the top left panel of each images.
%\label{fig:pictures2}}
%\end{figure*}

\begin{figure}
\includegraphics[width=0.45\textwidth]{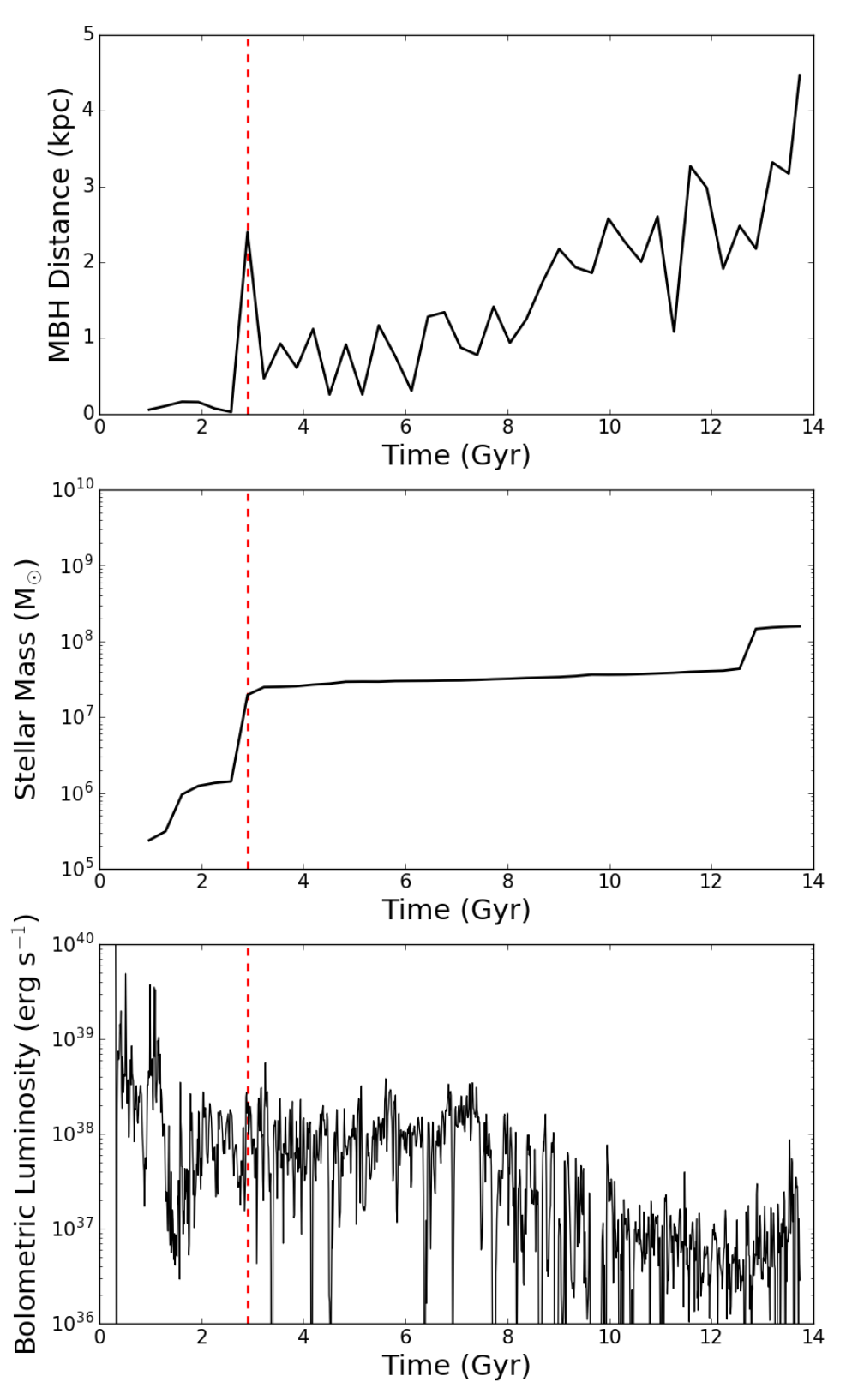}
\caption{{\em Top panel:} Distance of an MBH to the center of its host halo vs time.  It enters a much larger halo at $\sim 2.5$ Gyr, resulting in a sudden increase in distance to the center of its (new) host.  {\em Middle panel:}  Stellar mass of the MBH's host galaxy vs time.  The time of the galaxy merger is easily seen when the host mass rises suddenly.  {\em Bottom panel:} Bolometric accretion luminosity vs time of the MBH, binned in 10 Myr intervals.  The luminosity decreases as the MBH departs the central regions, due to lower gas densities.  In all panels, the red dashed line indicates the time the MBH leaves the center of its host halo.  \label{fig:storm1}}
\end{figure}

We show a graphical representation of this galaxy merger in the Storm simulation in Figure \ref{fig:storm1}. %and  \ref{fig:storm2}.  
We define ``merger'' as the moment when a satellite galaxy can no longer be distinguished by the halo finder (as opposed to the moment the satellite enters the virial radius (R$_{\rm vir}$) of the host, which happens earlier).  In this three-panel plot, the top panel shows the distance of the MBH from the center of its host galaxy.  The MBH in Figure  \ref{fig:storm1}  is the first to merge with the larger host, which occurs at the moment the distance suddenly jumps to a larger value (shown by a red dashed line in all panels).  This  jump occurs because the satellite host of the MBH has been disrupted, and the new host is the larger galaxy.   The center panel shows the stellar mass of the MBH's host galaxy as a function of time.  At the moment of the merger, the stellar mass also makes a sudden jump, which reflects the fact that the MBH has suddenly entered a much larger host galaxy.  The bottom panel shows the bolometric accretion luminosity of the MBH vs time, in bins of 10 Myr.  After an initial burst, the luminosity remains around $\sim 10^{38}$ erg s$^{-1}$ for several Gyr.  As the MBH moves further away from the galaxy center, however, the mean luminosity also decreases.  For more discussion regarding luminosity, refer to Section \ref{sect:lum}.

\begin{figure*}
\includegraphics[width=0.7\textwidth]{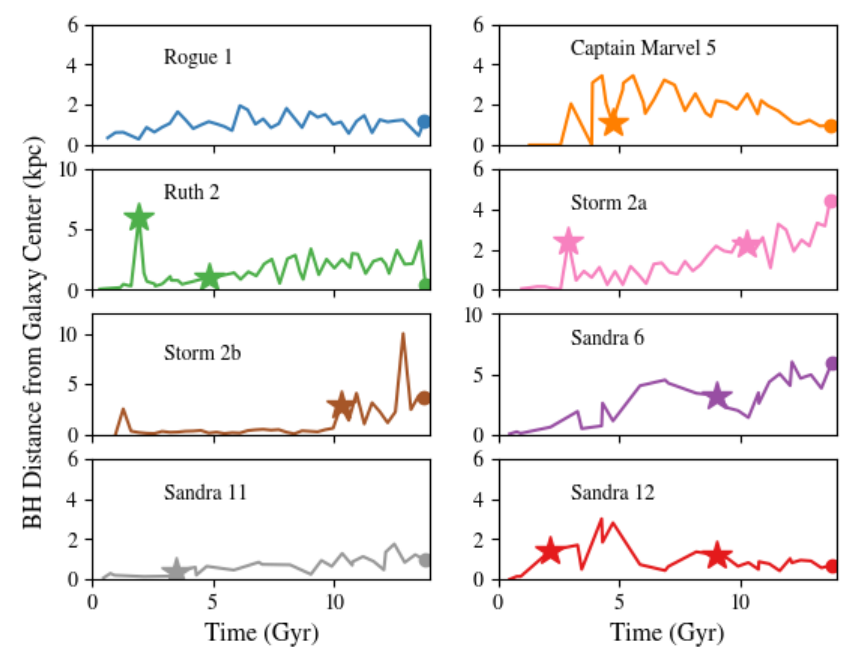}
\caption{Each panel represents a different wandering MBH, as shown by the labels.  Larger stars during MBH evolution represent galaxy mergers, which act to perturb the MBH from the center.  Circles at 13.8 Gyr  mark the final position for clarity.    \label{fig:distances}}
\end{figure*}

We show the wandering nature of all BHs together in Figure \ref{fig:distances}, where we plot the distance from the galaxy center vs time for each object.   There are no cases where a BH is dislodged from the center and eventually returns there for the long term,  nor are there cases where a central IMBH experiences a merger and remains central.  In fact, mergers at late times can cause MBHs to move even {\em further} from their hosts' centers.  In this Figure each MBH is depicted in a different colour and panel.  We place a coloured circle at the end of each line for clarity.  Larger stars at times between 0 and 13.8 Gyr represent galaxy mergers, which act to perturb the MBH from the center.  The galaxy Storm 2 experiences two such mergers (at 3 Gyr and 10 Gyr, the latter displacing the pink MBH a second time).  The galaxy Sandra 6 has a merger at 9 Gyr, but we note the MBH is already off-center at this time.  We suspect there was an earlier merger before 4 Gyr, but the time resolution of our simulation outputs for Sandra is poor, and we cannot confirm when this putative merger may have taken place.  These three MBHs are examples of distances {\em increasing} with time, as compared to the other five which remain approximately constant (while oscillating).  Our results are consistent with observations by \citet{Reines20}, in which the off-center compact radio sources are located between 1-5 kpc from the optical centers of their host galaxies.

Mergers with other galaxies are the cause of the displacement of MBHs in all but one case (Rogue 1, though possibly also Sandra 6).  In Table \ref{table:df} we show the properties of off-center MBHs and their host galaxies.  The column $z_{\rm off}$ represents the redshift at which the MBH becomes off-center when perturbed by a merger.    In all of these cases, the MBH is hosted by the ``minor'' galaxy in the merger, and coalesces with a more massive galaxy.  The resulting tidal stripping of the smaller galaxy leaves its remains orbiting in the halo of the primary galaxy.  The larger galaxy often lacks an MBH itself, which is a consequence of our MBH seed formation model.  To mimic the ``direct collapse'' scenario, we require gas to have a low H$_2$ fraction to form MBHs, while stars form with a high H$_2$ fraction.  As a result, gas in galaxies which form first (and are often larger as a result) will meet the criteria for star formation, but not MBH formation.  MBHs form from gas which has likely been heated by nearby star formation, consistent with the direct collapse model \citep{Spaans06}.

\begin{table*}
\begin{tabular}{lccccccc}
\hline
Simulation    & Halo  & $M_{\rm BH}$ & $z_{\rm off}$ &    $M_*$ &     $r_i$      &    $\sigma$    & $t_{\rm df}$  \\
Name & ID     & (\msun)                             &  &  (\msun)&   (kpc)    & (km s$^{-1}$)  & (Gyr)   \\
\hline
\hline
Sandra & 6 & $3.1 \times 10^5$ &  3.0 & $1.99 \times 10^8$  & 6.9 & 10.8 & 69.1\\
Sandra & 11 & $3.3 \times 10^5$ &  1.9   &  $8.12 \times 10^7$& 1.0 & 8.9 & 11.4\\
Sandra & 12 & $1.2 \times 10^6$ &  3.0    &  $6.26 \times 10^7$& 0.78 & 32.7 & 9.4\\
Ruth & 2 & $2.5 \times 10^5$ &  3.3  &  $4.50 \times 10^8$& 0.46 & 13.0 & 12.1 \\
Captain Marvel & 5 &   $2.4 \times 10^5$ & 1.6   &  $7.33 \times 10^6$  & 1.0  &  3.3  &  5.6    \\
Rogue &  1 &  $2.5 \times 10^5$&    8.1 &   $9.86 \times 10^8$&   2.4 &  47.1  &  158 \\
Storm &  2 &   $2.0 \times 10^5$ &   2.4  & $9.48 \times 10^7$  & 4.9  &  18.5  &  140 \\
Storm &  2 &  $7.6 \times 10^4$ &   0.3  &  $9.48 \times 10^7$ &3.8  &  18.5 &  297  \\

\end{tabular}
\caption{Off-center MBH and galaxy properties, including Halo ID at $z = 0$, black hole mass M$_{\rm BH}$, the redshift of offset $z_{\rm off}$, stellar mass M$_*$, distance to the galaxy center $r_i$, stellar velocity dispersion  $\sigma$ measured within $r_i$,  and dynamical friction timescale $t_{\rm df}$. Stellar masses are multiplied by an observational correction factor of 0.6 as described in \citet{Munshi13}.   \label{table:df}}
\end{table*}

\begin{figure}
\includegraphics[width=.45\textwidth]{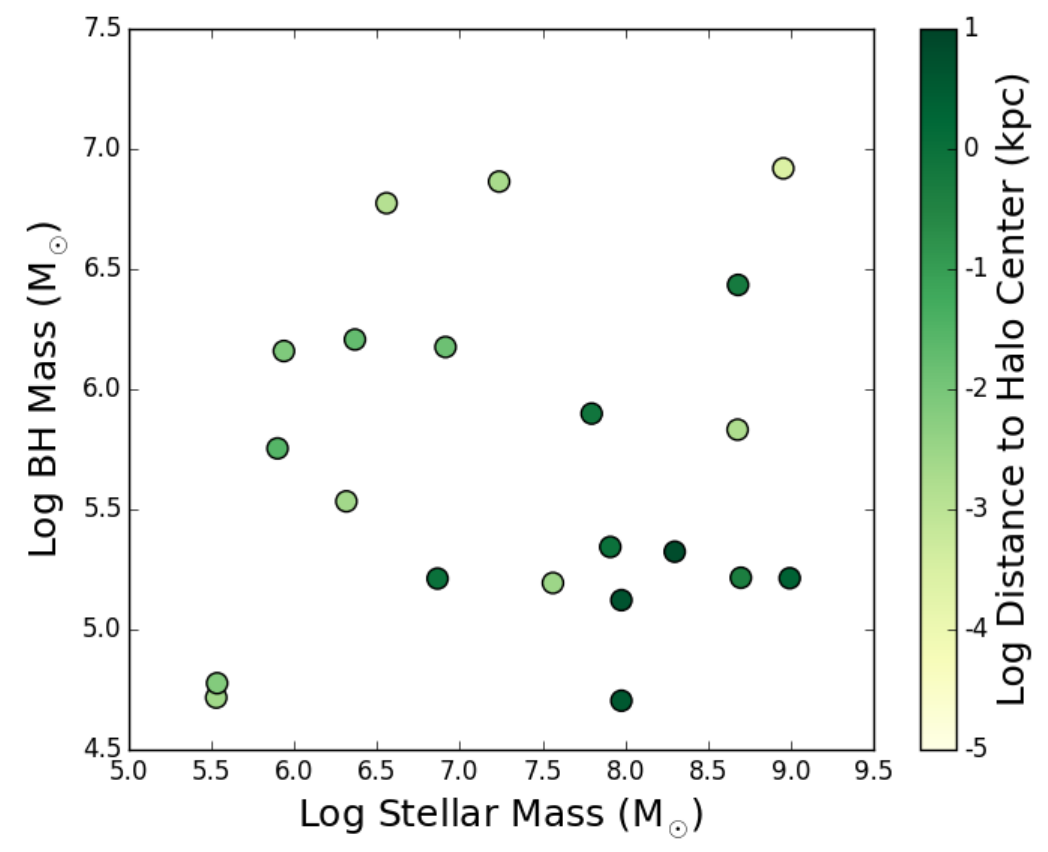}
\caption{Log MBH mass vs Log of the stellar mass for all dwarf galaxies hosting MBHs at $z = 0$.   Points are coloured by the log of the distance of the MBH to the galaxy center, with darker green points being more off-center.  Off-center MBHs are typically of lower mass but hosted in dwarfs with larger stellar masses.
\label{fig:stellarmass}}
\end{figure}

Overall, off-center MBHs are less massive and located in more massive hosts than those with central MBHs.  In Figure \ref{fig:stellarmass} we plot MBH mass vs host galaxy stellar mass (which is multiplied by 0.6 to correct for observational bias \citep{Munshi13}).  Points are coloured by the log of the distance to the halo center, with light circles representing central MBHs and darker green points representing further offset MBHs.  Dark green points are primarily located in the quadrant containing larger host stellar masses and lower MBH masses.  These instances occur when tiny MBH-hosting galaxies merge with larger (and often) MBH-less galaxies.  The larger hosts swallow the remains of the small galaxies, but the bulk of the disruption occurs in the outskirts of the larger galaxy.  These larger galaxies are still dwarfs; they do not have deep enough potential wells for dynamical friction to act to bring the MBH to the center.   The lower MBH masses also result in longer dynamical friction timescales, slowing their journey to the center as well  (see Section \ref{sect:why}).

The outlier in this story is the MBH in the largest galaxy of the Rogue simulation (known as Rogue 1).  The black hole forms around $z = 11$, but our first simulation snapshot is recorded 540 Myr later.   At that moment, the host galaxy is extremely bursty and messy, consisting of many clumps of localized star formation.  Due to our lack of time resolution at very high redshift, we cannot determine whether the MBH forms centrally and is quickly perturbed, or whether the MBH forms in an off-center clump.   Rogue 1 has a very bursty history overall and grows to become the largest dwarf galaxy in our sample (in Figure \ref{fig:profiles} it is represented by the topmost density profile).   This galaxy is an instance where extremely bursty star formation results in  likely off-center MBH {\em formation}, and dynamical friction is not effective enough to bring it to the center.

\section{Remaining Off-Center}\label{sect:why}

When MBHs in dwarf galaxies are perturbed from their centers, dynamical friction may be able to bring them back over time.  In massive and/or smooth galaxies this process is expected to be efficient, due to the dense stellar populations (e.g. galaxy bulges) surrounding SMBHs.  Dwarf galaxies have a variety of morphologies, but in general  they have a lower central stellar density than their massive counterparts, a clumpier structure, and a smaller scale radius as well.  Thus, when an MBH leaves the center, it is more likely to travel to a region of low density (possibly outside of the main stellar component of the galaxy) and has a much more tenuous stellar background to move through on its way back to the center.  As a result, dynamical friction timescales are quite long for these MBHs, which may take tens of billions of years to return to their galaxys' centers.  This phenomenon of non-sinking seeds is also explored in detail in \citet{Ma21}, who also find that dynamical friction timescales are longer than a Hubble time when MBHs are perturbed from the centers of clumpy, high-redshift galaxies.   

To calculate dynamical friction timescales, we employ the following equation from \citet{Binney08}:

\begin{equation}
t_{\rm df} \sim \left(\frac{19~  {\rm Gyr}}{{\rm ln} \Lambda}\right)\left(\frac{r_i}{5~ {\rm kpc}}\right)\left(\frac{\sigma}{200~ {\rm km/s}}\right)\left(\frac{10^8 {\rm M}_{\odot}}{ M_{\rm BH}}\right)
\end{equation}

\noindent
  $\Lambda$ is assumed to be equal to $b_{\rm min}/b_{\rm max}$, where b$_{\rm max}$ is the maximum impact parameter (which we set equal to the $z = 0$ radius of the MBH orbit, $r_i$), and $b_{\rm min}$ is the minimum impact parameter (set to 10 pc to represent the characteristic size of nuclear star clusters, which commonly exist around lower-mass SMBHs).  We directly measure $\sigma$, the velocity dispersion, of all of the star particles within $r_i$ in each galaxy.  To ensure we are making a conservative estimate of the DF timescales, we add a factor of 50 per cent to the mass of each MBH, to account for the additional mass of a possible nuclear star cluster, which will shorten the DF timescale estimate.     These timescales are not affected by the low densities in galaxy outskirts, because the subgrid model interpolates the unresolved density.  Even for a higher resolution simulation with more star particles, the unresolved density would not be affected as long as the background density of stellar (and dark matter) mass is the same.  The dynamical friction model has been tested at a range of resolutions and is found to match the Chandrasekhar estimate well at the resolutions we use here, even in cases of low particle density.  
  
In Table \ref{table:df} we present information for each off-center MBH.  We acknowledge that as the distance of each MBH to its galaxy center changes, the values of $t_{\rm df}$ will vary, and thus using $r_i$ at the present time ($z = 0$) gives a basic estimate of this timescale.  Even as an order-of-magnitude estimate, however, one can see that the majority of the timescales are longer than a Hubble time, which does not change when using a different radius.  For the majority of these off-center MBHs, they will effectively never return to the centers of their hosts.

\begin{figure}
\includegraphics[width=0.45\textwidth]{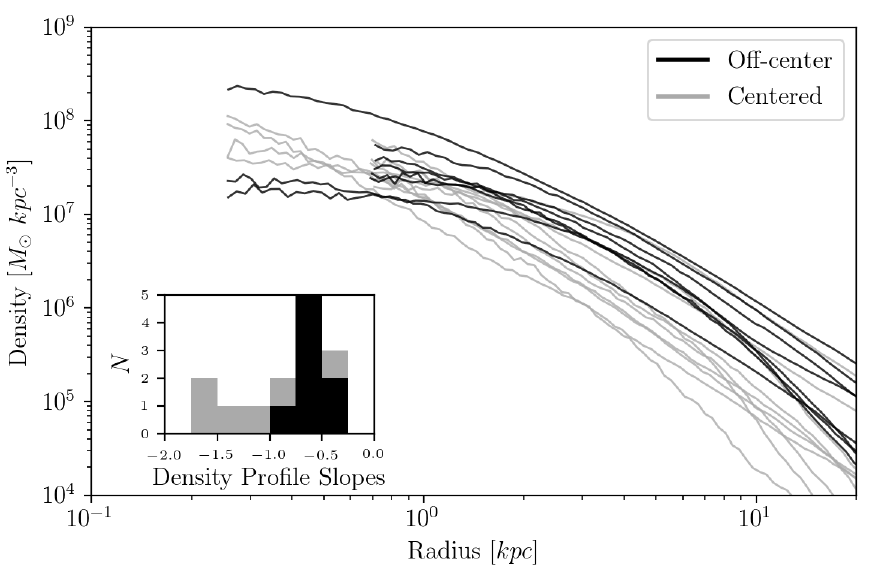}
\caption{Density profiles of all MBH-hosting dwarf galaxies in our sample.  Dark lines represent galaxies with off-center MBHs, and grey lines represent those with central MBHs.  The inset shows a histogram of the slopes of the inner density profiles, with the black histogram representing galaxies hosting off-center MBHs.  All of the galaxies with off-center MBHs exhibit cored profiles.  \label{fig:profiles}}
\end{figure}

 The shape of the potential well in dwarf galaxies is another reason why their MBHs may remain off-center; defining a ``center'' is not always trivial.  For example, the Large Magellanic Cloud has been reported to have several conflicting measured centers (e.g. HI kinematics, stellar kinematics, and photometry each give a different answer \citep{vandermarel14}).  Dwarf galaxies are often irregularly shaped, and also often exhibit cored density profiles \citep{Moore94,Relatores19}.  This is the case in our simulated galaxies as well, as shown in Figure \ref{fig:profiles}.  Dark lines represent galaxies which host off-center MBHs, and grey lines represent those with central MBHs.  The distribution of inner slopes is shown in the inset, with the grey depicting all galaxies and the black region galaxies hosting off-center MBHs.   We fit the dark matter profiles using the core-Einasto profile presented in \citet{Lazar20}.  To derive the inner slope, we fit from a line to the profile from our innermost resolved radius (set conservatively to 4 times the gravitational softening) out to 1-2\% of the R$_{\rm vir}$.  We use the slope of this line ($\alpha$) as the core slope we present in Figure \ref{fig:profiles}. Slopes steeper than -1 are considered to be cuspy, whereas slopes that are less than -1 to 0 are showing core formation, or large cores. 
 
All galaxies hosting off-center MBHs have central densities with flat slopes, indicating cored profiles; galaxies with central MBHs have a combination of cored and cuspy profiles.   Past simulation work has demonstrated that  cored profiles in dwarfs have been shown to be a result of a redistribution of mass by driven by supernova feedback \citep[e.g.][]{Governato12}.  Dwarfs with cuspy profiles likely have less bursty star formation histories.   Due to the shallow shapes of cored profiles, a displaced MBH does not feel a strong gravitational pull to the galaxy center.  Instead, it may experience core-stalling, ``sloshing'' around within the inner region (typically 1-2 kpc) without settling into one location.    50\% of the off-center MBHs in our sample are located within the approximate ``core'' region of their galaxies, and all of those have existed within a mean radius of 2 kpc or less for the majority of their off-center lifetimes, indicating approximate core-stalling behaviour.  (The dynamical friction subgrid model prevents a literal ``stall'' because it continually acts, but the timescales are still very long, effectively stalling the orbital decay.)   In fact, only those galaxies which experience recent mergers have MBHs which are outside of the central core.  (see Section \ref{sect:how}).

One might ponder how these results might change with a different seed model, e.g. ``light'' seeds (from Population III stars)  plus efficient early growth.    With such a model, seed formation is much more efficient, and we might expect every dwarf galaxy to host (at least one) MBH.  The masses of such seeds is poorly constrained, but we can conjecture that they may be around $\sim 1000$ \msun.  These seeds are not constrained to form in galaxy centers, and so their initial locations  within galaxies would vary.  Due to their lighter masses, they would experience even weaker dynamical friction forces compared to their ``heavy'' counterparts.  Thus, one might expect an even greater number of wandering MBHs in dwarf galaxies, whether or not they have undergone galaxy mergers, since the efficiency of orbital decay to the central region decreases.  On the other hand, with a greater number comes a greater statistical likelihood of an MBH undergoing something more interesting, like settling into the center, or increasing its mass rapidly (perhaps by undergoing a large accretion event or merging with another MBH).   Overall we postulate that light seeds would result in an increased number of (lower mass) wandering MBHs, but a buildup of a more massive central MBH can't be ruled out.  Cosmological simulations with such small particles run to $z = 0$ are extremely computationally intensive, but this topic would be worthwhile to pursue in future work.

\section{Detecting Off-Center MBHs}\label{sect:detect}

Our work shows that off-center MBHs in dwarfs maybe be fairly common, especially for those with larger stellar masses, due to the combination of a higher MBH occupation fraction \citep{Bellovary19} and higher off-center incidence with increasing M$_*$.  Recently, observations have uncovered evidence for these objects as well.  In \citet{Reines20}, they examined radio-selected dwarf galaxies and found 13 with clear signatures of compact radio sources due to AGN, and approximately half of these are off-center.  Subsequently, \citet{Mezcua20} used spatially resolved emission line diagnostics to detect AGN signatures in 37 dwarf galaxies.  They report that in most cases the AGN emission is offset from the photometric center of the galaxy, possibly indicating off-center MBHs.  In this section we address the detectability of our simulated off-center MBHs, both electromagnetically and via dynamical signatures.

\subsection{Luminosity}\label{sect:lum}

We calculate a bolometric luminosity due to accretion onto the MBH using the formula $L = \epsilon_f \epsilon_r \dot M c^2$, where the accretion rate $\dot M$ is described by Equation \ref{eqn:mdot} and  is  dependent on the density of the surrounding gas.  Gas in galaxy halos tends to be quite diffuse, resulting in relatively low MBH accretion rates compared to galaxy centers.  The accretion rate also depends on the relative velocity between the MBH and the surrounding gas ($v_{\rm bulk}$), which is much larger for an MBH drifting through its host galaxy compared to one residing stationary at the center.   It is also worth noting that the MBHs in dwarfs have categorically  lower masses compared to SMBHs in galaxy centers.  As a result of the lower densities, higher velocities, and lower black hole masses, substantial accretion rates are not expected.

\begin{figure*}
\includegraphics[width=6in]{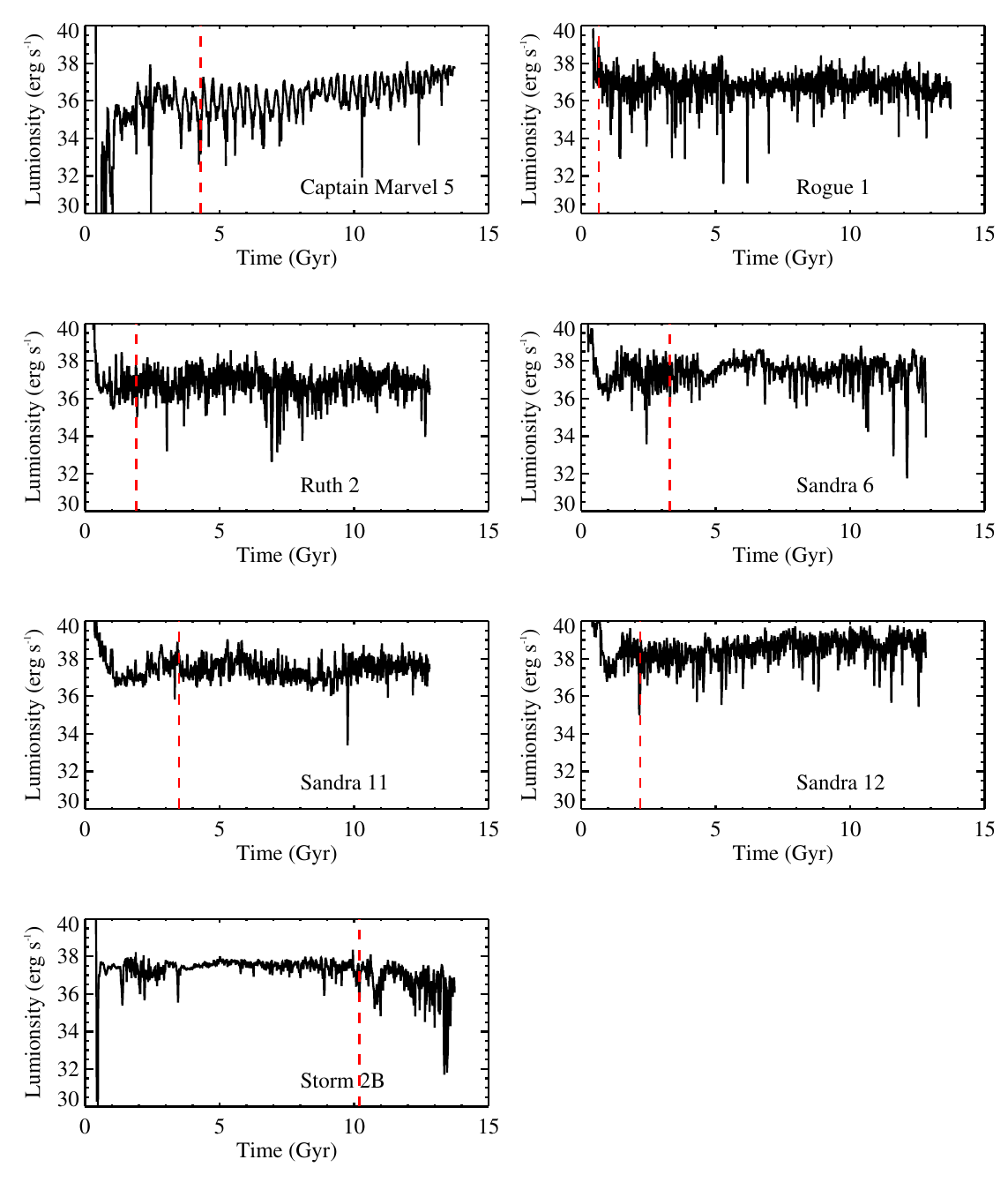}
\caption{ Bolometric luminosity vs time for every MBH which resides off-center in a dwarf galaxy at $z = 0$.  Luminosities are binned in 10 Myr intervals.  The simulation name and halo ID are labeled on each plot.  All MBHs are below detectability thresholds at all times, and luminosities either are unchanged or decrease once they become off-center.  The red dashed vertical lines indicate the moment the MBHs become off-center.  \label{fig:lightcurves}}
\end{figure*}

Figure \ref{fig:lightcurves} shows light curves for every off-center MBH in a dwarf from the time of formation to $z = 0$ (excepting the Storm 2A MBH light curve which is shown in Figure \ref{fig:storm1}).  The luminosities are binned in 10 Myr intervals. We point out that the y-axis shows bolometric luminosity, so  the luminosity in any given part of the electromagnetic spectrum (e.g. X-rays) will be at least 10\% of this value.   Each curve is highest just after the MBH forms, which is an artifact of MBH formation in {\sc ChaNGa}.  Because seeds form from cold dense gas, they undergo a short accretion event before their accretion feedback can heat the gas.  Once feedback occurs, the surrounding gas quickly reaches an equilibrium in density and temperature.  

All of the light curves vary dramatically, but maintain average values around $10^{36} - 10^{38}$ erg s$^{-1}$.  At the moment they become off-center (shown by the dashed red line), some of the mean luminosities tend to decrease, due to the decrease in ambient gas densities and increase in $v_{\rm bulk}$.  Other MBHs undergo little luminosity change at all.   Accounting for bolometric corrections, none of these simulated off-center MBHs can be easily distinguished as such by electromagnetic radiation at any point in their histories.  

While this result is an apparent contradiction to observations by \citet{Reines20} and \citet{Mezcua20}, we point out that our sample size is quite small.    These and other works report AGN fractions in dwarfs of a few percent or less \citep{Reines13,Pardo16,Mezcua18}, suggesting that the conditions for MBHs to exist as AGN in dwarfs are quite rare.  We also point out that our calculated luminosities depend on a modified Bondi-Hoyle accretion model, which may not be an accurate representation of the physical situation.  Additionally, several of our MBH host galaxies have stellar masses less than M$_* < 10^8$\msun, and there are very few reliable detections of AGN in such small galaxies.   If we add the complications of wandering to this situation, it is not surprising that off-center MBHs with high accretion rates in dwarfs are extremely rare.  Rather, it is perhaps surprising that off-center MBHs have been detected electromagnetically in dwarfs at all.

The existence of a nuclear star cluster could improve the situation, however.  Lower-mass SMBHs tend to have nuclear star clusters \citep{Boker10}, which may remain bound to a perturbed MBH.  The luminosity of the stars in the cluster would increase the likelihood of detecting an off-nuclear point source.  Additionally, if one of the bound stars becomes dynamically perturbed, it may interact with the MBH and create a tidal disruption event or similar hyperluminous  X-ray source.  The object HLX-1 \citep{Farrell09} is an example of such a candidate, and is consistent with a dwarf galaxy which hosts an MBH disrupting an orbiting star \citep{Lasota11}.     While we cannot resolve nuclear star clusters or tidal disruption events, it is worth keeping these scenarios in mind when discussing observability.

\subsection{Dynamics}\label{sect:dynamics}

While MBHs are not observable due to their accretion luminosity, their effects may be discernible dynamically, via their interactions with stars.  A massive object may ``stir''  stars in the galaxy, creating an increased velocity dispersion, altered anisotropy profile, or other noticeable signature.  Such an effect is not measurable in our cosmological simulations, which do not resolve the radius of influence of the MBH.  We have thus followed up on this concept by further ``zooming'' in and resimulating dwarfs with off-center MBHs at higher resolution.

The galaxy we selected for further study is the fifth most massive galaxy in the Captain Marvel simulation (hereafter the fiducial simulation); it has a stellar mass of $1.2 \times 10^7$ \msun ~and a total halo mass of $8.6 \times 10^9$ \msun.  The MBH is 1.0 kpc from the center at $z = 0$ and has a mass of $1.6 \times 10^5$ \msun.   This halo has a benign history since its last major merger at $z \sim 1.4$, which initially caused the off-center MBH, and exists in isolation.  We made a spherical cutout of the $z = 0$ galaxy at the virial radius, and increased the particle number by a factor of 8.  The resulting galaxy has over 11 million total particles, with masses $M_{\rm gas} = 70$\msun, $M_{\rm star} = 32$\msun, and $M_{\rm dark} = 832$\msun.   Star and dark matter particles were split by placing new particles along a 3D Gaussian kernel centered on the parent particle, while for gas particles we replaced the particle with 8 particles at the corners of a cube of size 0.5 the mean interparticle separation, and then rotated that cube to a random orientation.   The MBH particle was not split.  The gravitational softening values of particles were decreased substantially to 0.5 pc, in order to resolve the sphere of influence of the MBH.  We ran this simulated galaxy in isolation for two billion years, as well as two additional simulations;  an identical one without an MBH (hereafter known as ``nobh''), and an identical one but with the MBH mass increased by a factor of 10 (referred to as ``heavy'').   Comparing these three simulations allows us to determine whether the effect of the MBH is noticeable compared to a galaxy without one, as well as to the case where the galaxy hosts an unreasonably large MBH (10\% the mass of the entire stellar population).

\begin{figure*}
\includegraphics[width=2in]{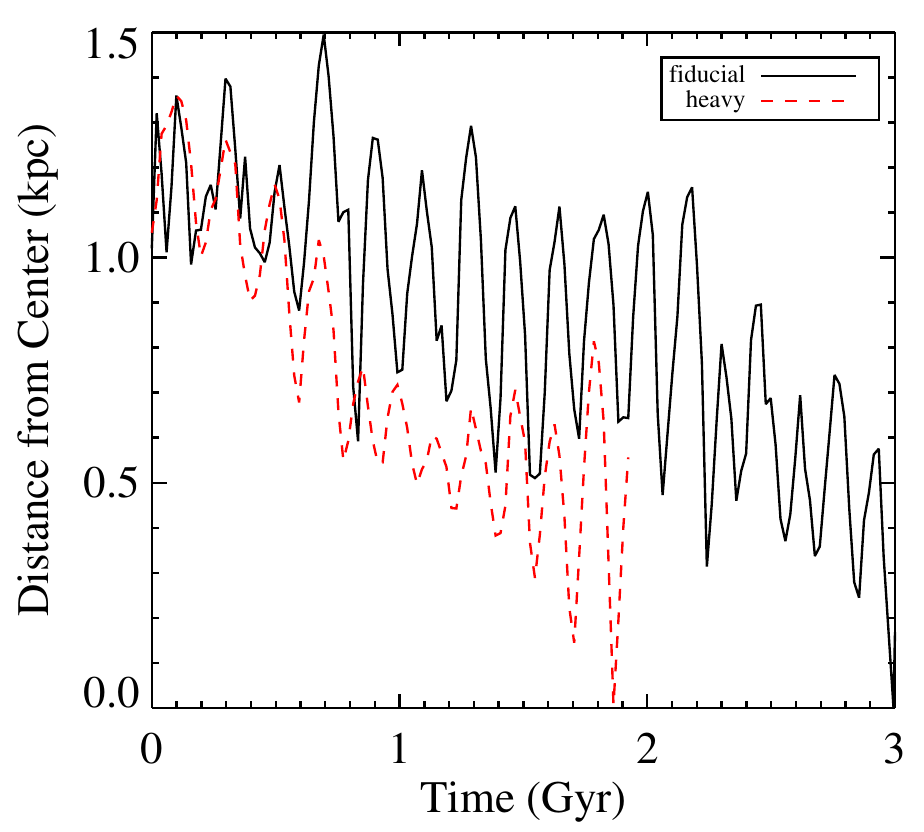}
\includegraphics[width=2in]{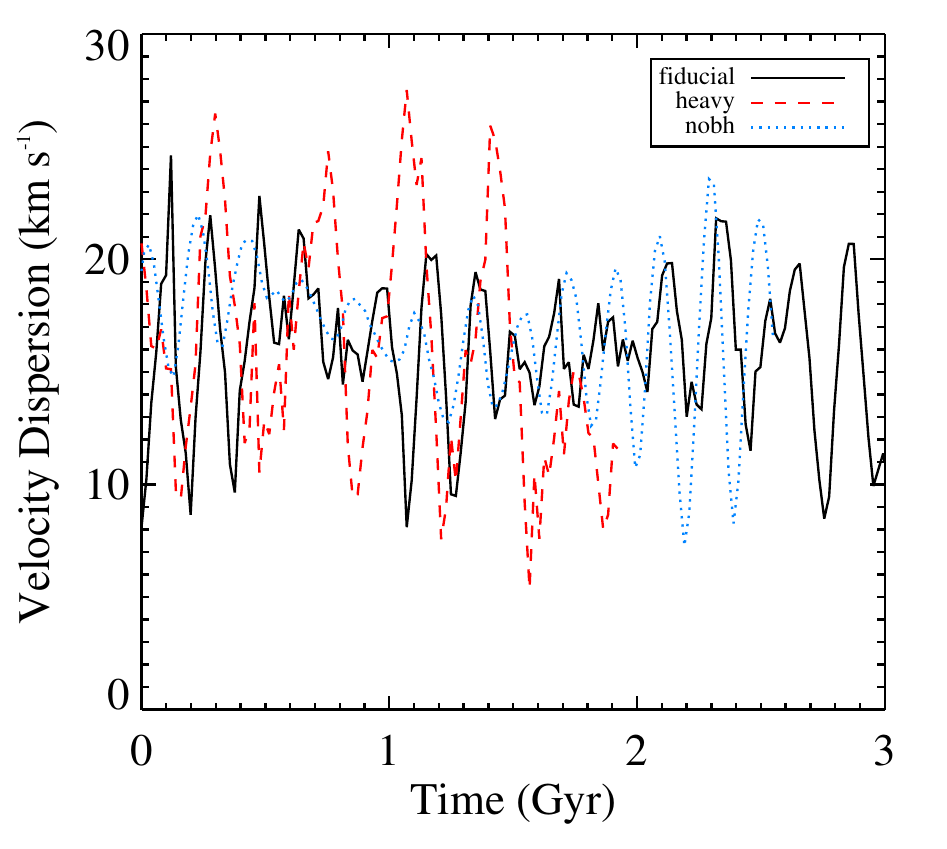}
\includegraphics[width=2in]{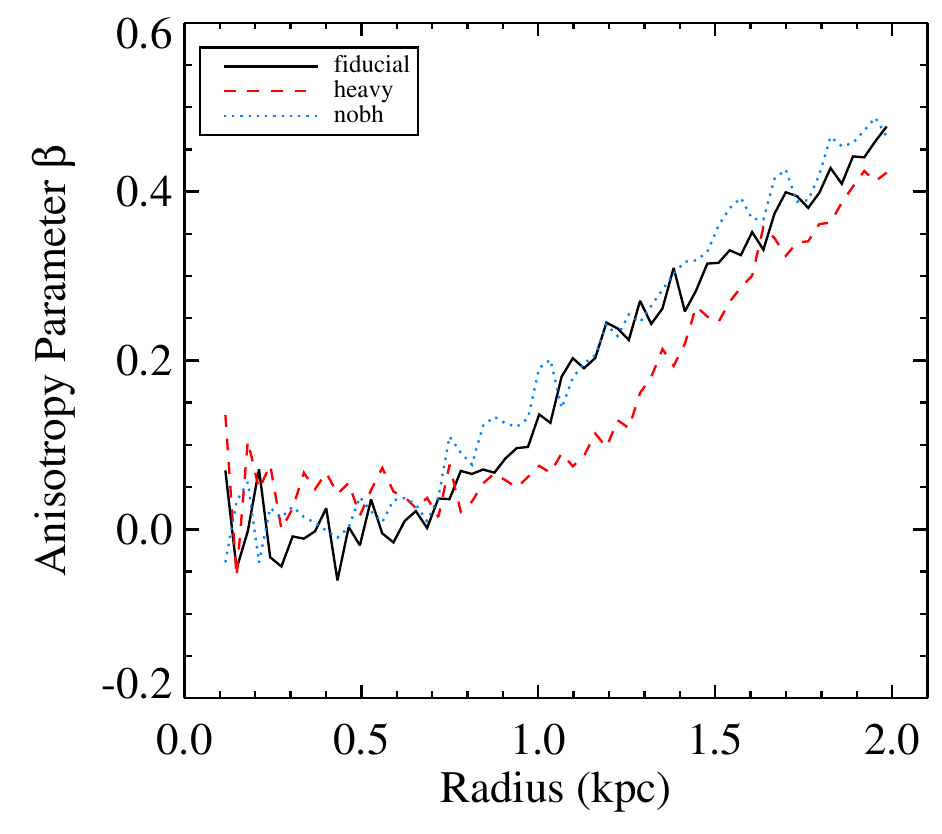}
\caption{{\em Left:}  The distance of the MBH from the galaxy center vs time.  {\em Center:} The velocity dispersion of the stars in the galaxy  vs time.  {\em Right:} Anisotropy profiles of each galaxy after 2 Gyr of evolution.  In all figures, the black solid line represents the fiducial run, red dashed line is the heavy BH run, and the blue dotted line is the run with no MBH.   \label{fig:katty}}
\end{figure*}

 Our goal with this experiment is to resolve the effects of both dynamical friction and two-body relaxation directly.  The black hole is far more massive than the surrounding particles, and thus will exchange energy with stars during gravitational interactions on small scales.  As a result, the MBH will lose energy and sink to the center of the potential well, while star particles will gain energy as their velocities are  increased.  We aim to determine if this increase in  stellar velocity is measurable, compared to a galaxy which does not host a wandering MBH.

In the left panel of  Figure \ref{fig:katty}, we show the distance of the MBH from the galaxy center vs time.  The heavier MBH, represented by the red dashed line, experiences stronger dynamical friction forces and sinks to the center more quickly, but does not quite get there within 2 Gyr.  The lighter MBH, shown with the black solid line, does experience dynamical friction but not enough for it to settle into the center within the following 3 Gyr.  These timescales are consistent with the Chandrasekhar dynamical friction estimate mentioned in Table \ref{table:df}, i.e. we calculated $t_{\rm df} = 5.6$ Gyr for the fiducial case studied in this section.

One might expect an increased velocity dispersion in a galaxy with an MBH (and a larger increase with a larger MBH), because gravitational interactions between the MBH and close-passing stars will cause stars to accelerate, thus increasing the velocity dispersion.   The dynamical situation is complicated, however, by the ``breathing'' nature of this galaxy, where episodic star formation causes semi-periodic expansion and contraction of the galaxy \citep[see][]{Stinson07}.    Comparing the velocity dispersion of the stars over time, the central panel of Figure \ref{fig:katty} shows that while the total dispersion varies over time in an oscillatory way, the minimum and maximum values are approximately equal, and no discernible difference exists.   Even in the case of the heavy MBH, the extrema of the values of the velocity dispersion are a bit larger than in the other cases, but not different enough for an observer to measure an anomalous dynamical state.   We suspect that the lack of difference is because there are few stars within the MBH's radius of influence at any given time.  

 Calculating the radius of influence ($R_{\rm infl} = GM_{\rm BH} / \sigma^2$) for each MBH, we find 2.7 pc for the fiducial case and 38 pc for the heavy case.  (Note that the force softening is 0.5 pc in both cases.)  These estimates of $R_{\rm infl}$ are when the MBH is in the denser regions of the galaxy; however, each MBH spends substantial time at the galaxy outskirts, when there are few or no star particles within $R_{\rm infl}$.  The median number of star particles within $R_{\rm infl}$ varies with the MBH's position, but in denser regions comes to $\sim 1-2$ for the fiducial  case and $\sim 40$ for the heavy case.  We also emphasize that being within $R_{\rm infl}$ is not identical to being bound, because both the MBH and stars are moving on their own trajectories.  The stars within $R_{\rm infl}$ at one moment are different from those within $R_{\rm infl}$ the next.   We propose a modified calculation for $R_{\rm infl}$ for a moving MBH:  $R_{\rm infl, moving} = GM_{\rm BH} / (\sigma^2 + v_{\rm BH}^2)$, which takes into account the movement of the MBH and effectively reduces the radius of influence.   Using this modified value, we recalculate $R_{\rm infl, moving} $= 1.3 pc for the fiducial case and 15 pc for the heavy case, with median value of 0 and 15 enclosed particles within each, respectively.  These lower numbers indicate that a moving MBH has a much lower chance of gravitationally influencing nearby stars.  Overall, this dwarf galaxy has a much lower stellar density than e.g. a galaxy bulge or nuclear star cluster, and so the MBH's gravitational influence is negligible on the stellar dynamics.  We have also investigated the potential effect  of the MBH on the dynamics and structure of each galaxy's dark matter and gas properties, and find no measurable differences.

Velocity anisotropy is a dynamical tool which has been used to infer the presence of (albeit central and unmoving) SMBHs \citep[e.g.][]{Binney82,vandermarel94}.  This quantity is a function of the velocity dispersion in the tangential direction $\sigma_\theta$ and the velocity dispersion in the radial direction $\sigma_r$, and is defined as 

\begin{equation}
\beta = 1 - \sigma_\theta^2 /2 \sigma_r^2
\end{equation}

\noindent
When measured in spheroidal galaxies, the velocity anisotropy typically ranges from $\beta = 0$ at galaxy centers (where stellar orbits are predominantly isotropic) to $\beta = 1$ at galaxy outskirts (where radial orbits dominate).  We measure the anisotropy profiles of our three models after 2 Gyr of independent evolution (right panel of Figure \ref{fig:katty}). The anisotropy profiles of the stellar populations are effectively identical in all three cases, which is not surprising since the MBH is not stationary, and the stars do not have time to adjust their orbits to its gravitational potential.  

\begin{figure}
\includegraphics[width=0.4\textwidth]{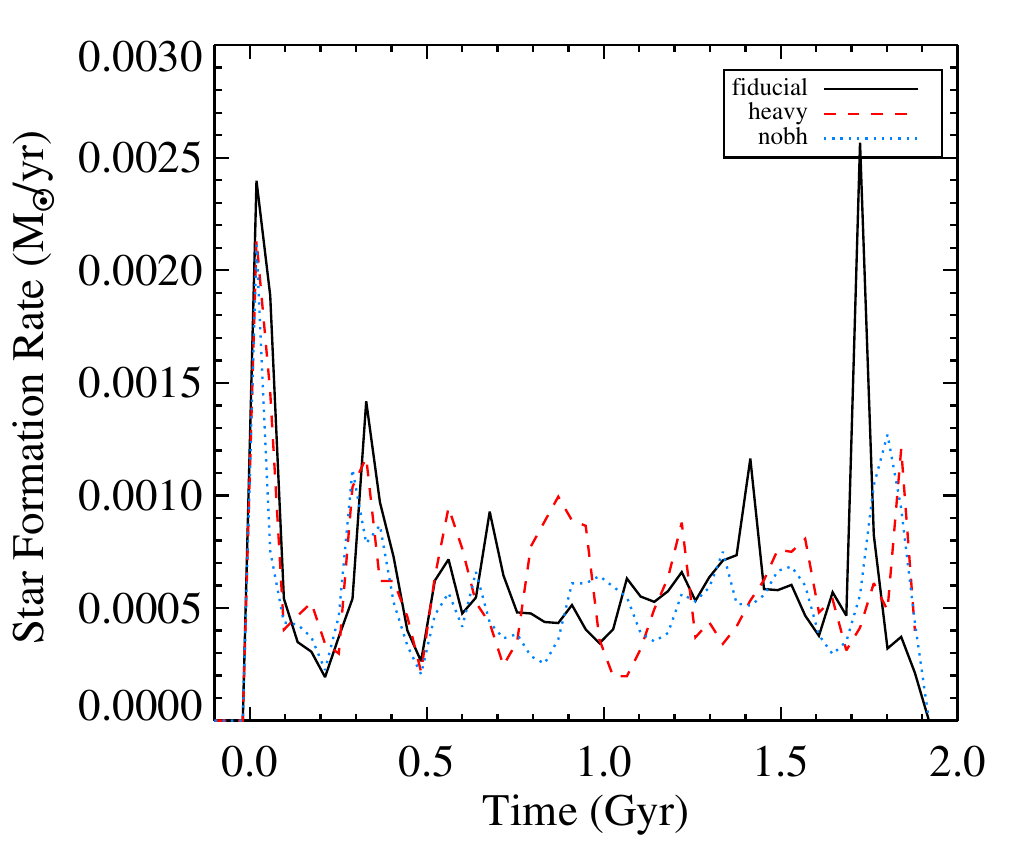}
\caption{Star formation history for the fiducial, heavy, and no MBH runs.  Lines are as in Figure \ref{fig:katty}.  \label{fig:sfh}}
\end{figure}

The star formation histories of the three cases are also extremely similar.  In Figure \ref{fig:sfh}, we show the star formation rate vs time for the three simulations, which begin nearly identically and then diverge.  However, the rates do not deviate from each other significantly.  One might expect additional quenching due to the presence of the MBH, but its movement prevents the accretion of substantial gas and thus any feedback effects.  One might also expect the passage of the MBH to trigger bursts of star formation by perturbing gas clouds, but this effect is not seen either.

\begin{figure}
\includegraphics[width=0.4\textwidth]{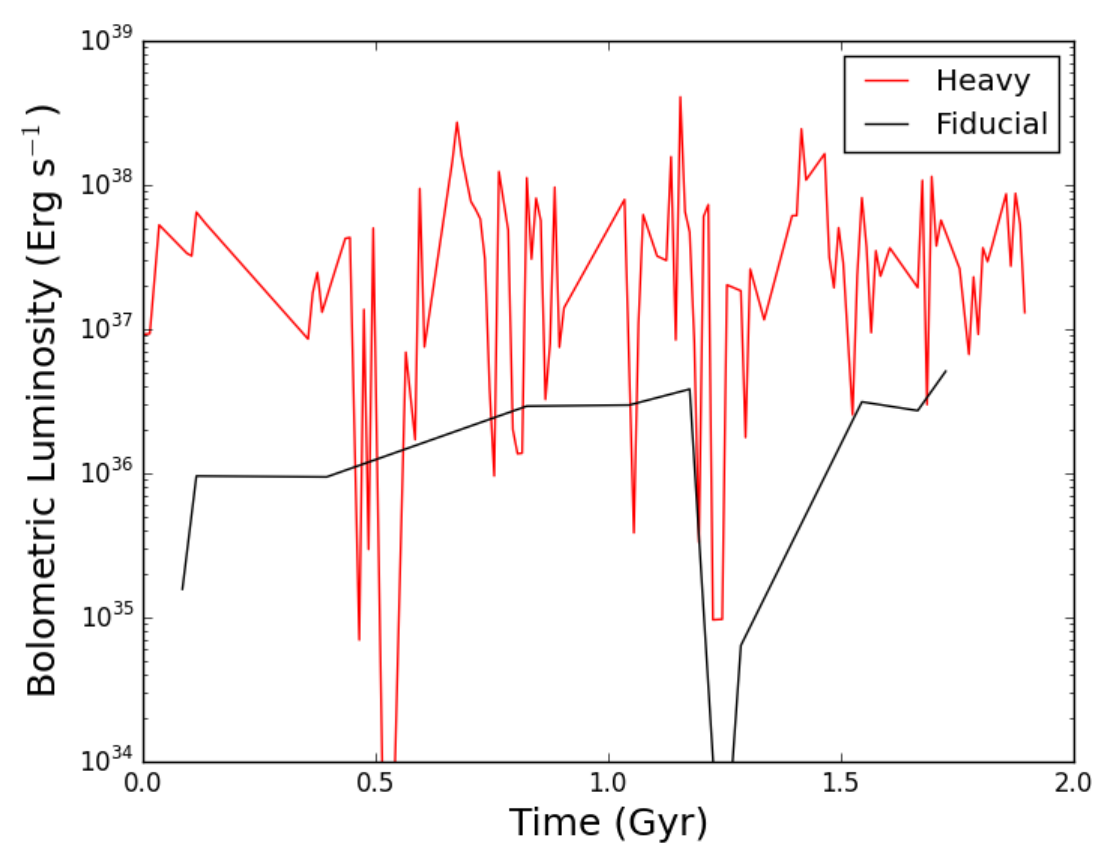}
\caption{Bolometric luminosities for the wandering MBHs in the isolated simulations, averaged over intervals of 10 Myr.  The black line is the fiducial simulation, while the red line is the heavy black hole.   \label{fig:lums}}
\end{figure}

The accretion luminosity of the MBH is not high enough to be detectable at any point.  In Figure \ref{fig:lums}, we show the bolometric accretion luminosity vs time for both simulations with MBHs.  Due to the wandering nature of the black holes and the diffuse gas in the galaxy, there is little opportunity to accrete a substantial amount of gas at any given time.   The heavier MBH has a larger luminosity, which is expected due to the nature of our accretion model ($\dot M \sim M_{\rm BH}^2$).  However, since both MBHs have a high relative velocity with respect to the gas, the accretion rate remains low.  Assuming a simple bolometric correction factor of 10\% for X-rays in the 2-10 keV range, an object with $L_X = 10^{37}$ erg s$^{-1}$ would not be discernible from an X-ray binary system, especially at an off-center location in the galaxy.

\begin{figure}
\includegraphics[width=0.4\textwidth]{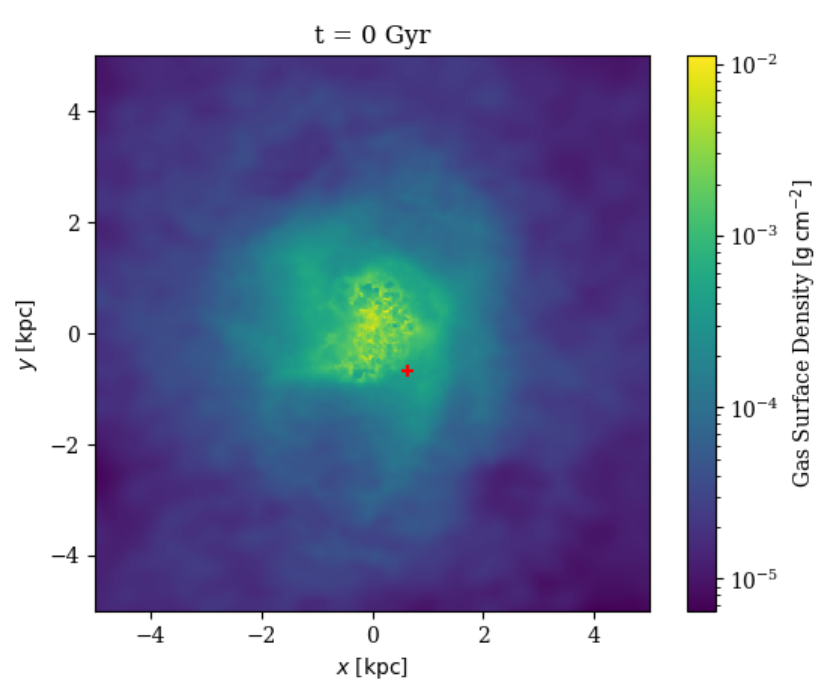}
\includegraphics[width=0.4\textwidth]{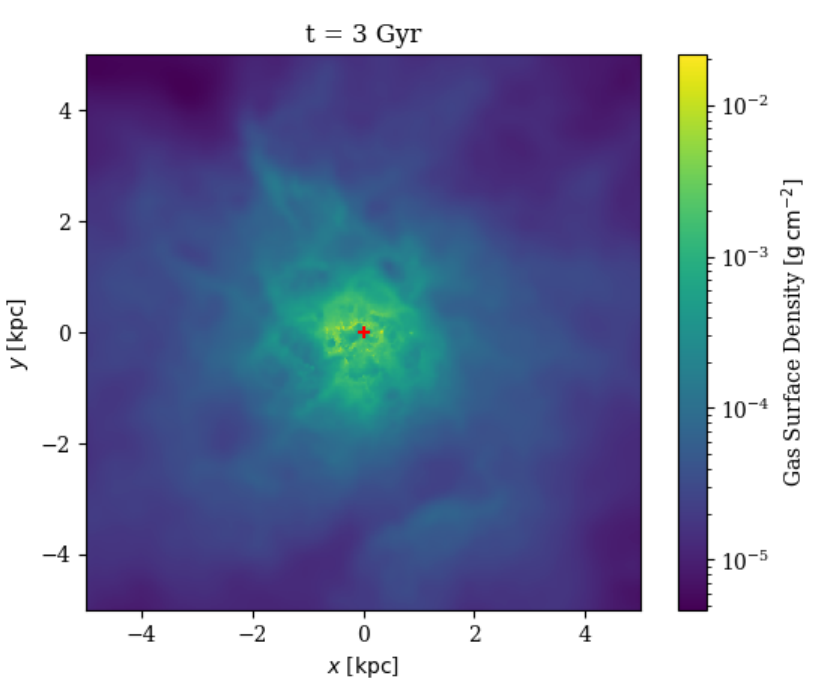}
\caption{ Images of projected (two-dimensional) gas density for the fiducial galaxy at $t = 0$ Gyr (top) and $t = 3$ Gyr (bottom).  The MBH is marked with a red cross.  The color bar indicates gas density.  The MBHs do not experience substantial accretion events because the surrounding gas is diffuse. \label{fig:gas}}
\end{figure}

In a more physically realistic accretion scenario, the gas must either become gravitationally bound to or experience a rapid inflow toward the MBH in order to be accreted, which would still be difficult in a case where the MBH is moving quickly through a diffuse medium.  Such a scenario is only likely in a galaxy with a larger number of clumps of dense gas (as seen in \citet{Reines20}), which would need to become bound to the MBH as it passes nearby.  In Figure \ref{fig:gas} we show gas surface density images of the fiducial isolated galaxy, at $t = 0$ (effectively at $z = 0$) and at $t = 3$ Gyr (3 Gyr later).  At $t = 0$, the MBH is moving through a region of lower gas density.  As a result, there is only low-level accretion from passing through this smooth gas.   At $t = 3$ Gyr, the MBH arrives at the galaxy center.  However, the gas distribution is irregular, and there are no clumps which could bind to the MBH.  Our simulations are capable of resolving a clumpy interstellar medium, and so the lack thereof is not a resolution issue.  Without a clumpy medium, an  overdensity, or a rapid inflow of gas, the MBH accretion will remain fairly quiescent.

We calculate the $B - V$ colours of each galaxy using the pynbody analysis suite, which uses simple stellar populations models  \citep{Marigo08,Girardi10} to convert stellar ages and metallicities to luminosities and subsequently magnitudes.  Including all stars in the galaxy in our analysis, the resulting colours are  $(B - V)_{\rm no BH} = 0.507$, $(B - V)_{\rm fiducial} = 0.527$, and $(B - V)_{\rm heavy} = 0.535$.  While the MBH-hosting galaxies are technically redder than their MBH-less counterpart, the difference is minuscule, and not detectable.

%  B-V (fiducial) =  0.527146597907
%  B-V (fatso) =  0.535123853314
%  B-V (nobh) =  0.506800208173

\section{Summary}\label{sect:summary}

Consistent with recent observations \citep{Reines20}, zoom-in cosmological hydrodynamical simulations show that about 50\% of MBHs in dwarf galaxies  are not located in the centers of their hosts.  In this paper, we examine the dynamical histories of these MBHs and analyze the likelihood of detecting them.

MBHs form in very low-mass halos ($10^8-10^9$\msun) and nearly always form centrally.  These MBH host galaxies undergo mergers with more massive dwarfs (often {\em not} hosting an MBH) during their evolution.  This merger perturbs the matter in both galaxies, but often results in the MBH host being tidally  disrupted by the larger galaxy.  The remains of the original MBH host thus join the components of the larger 
dwarf's stellar halo, where the MBHs are seen wandering.  However in one case out of eight, the black hole does form off-center and remains so, and thus it is possible that a small fraction ($\sim 10$\%) of off-center MBHs are not due to perturbations from mergers.  Indeed, factors other than those described here may also contribute to wandering MBHs in dwarfs, such as perturbations from merging subhalos \citep{Boldrini20} or gravitational recoil \citep[e.g.][]{Blecha11}.

The combination of dwarfs exhibiting cored density profiles and low stellar densities results in long dynamical friction timescales.  As a result, the MBHs which enter galaxies via mergers do not sink to the center within a Hubble time.  The most likely off-center MBH hosts are galaxies with larger stellar masses (i.e. more likely to have acquired a smaller galaxy in a merger) and smaller MBH masses (resulting in weaker dynamical friction forces).

These off-center MBHs are extremely difficult to detect.  Their accretion rates are very low due to the low density of gas in galaxy halos and the high relative velocities of the MBHs themselves.  The accretion luminosities of these objects are low enough to either be undetectable or easily confused with another object, such as an X-ray binary or background quasar.  We also examined whether an off-center MBH could be detected indirectly via dynamical studies of the surrounding stars.  A massive compact object might dynamically heat the stellar system, resulting in larger velocity signatures.  Unfortunately, due to the low masses of the MBHs and the low stellar densities, there are too few stars within the radius of influence of the MBH at any given time to cause any difference in dynamical signatures.

One detection method we do not explore in this work is strong gravitational lensing.  \citet{Banik19} predict that wandering IMBHs  can distort the lensing arcs formed by dark matter haloes when they cross the line-of-sight either in the foreground or background  of the lens.  Such distortions can be detected by high resolution interferometers such as the SKA.  Alternatively, \citep{Paynter21} report a gamma-ray burst which may be lensed by an IMBH, based on a delay in the arrival pulse.  We encourage further investigation of this fascinating detection method, which is beyond the scope of this paper.

While the lack of detectability of off-center MBHs is disappointing for those who seek to find more of them, we remind the reader that our sample size is very small.  The AGN fraction of dwarf galaxies is already quite low ($<1\%$ \citep{Reines13,Pardo16,Mezcua18}), and this estimate is for central MBHs.  Simply based on statistics., one does not expect to find a luminous AGN  in a sample as small as ours.  For a larger sample, in future work we will look to the {\sc Romulus} simulation \citep{Tremmel17}, which hosts 492 dwarf galaxies (defined as isolated and with  stellar masses $10^8 < M_* < 10^{10}$\msun); 228 of these host MBHs \citep{Sharma20}.  Repeating our analysis on this larger sample will provide new insights to the causes and detectability of off-center MBHs.

\section*{Data Availability Statement}
The simulations analyzed in this work (DC Justice League and MARVEL-ous Dwarfs) are proprietary and are not available to the public.  The authors are happy to share quantitative data related to our results  for collaborative purposes upon request.

\section*{Acknowledgements}

The authors thank the anonymous referee, who asked helpful and interesting questions.  JMB is grateful for the support of the AstroCom NYC project (NSF AST-1153335).  JMB and SH acknowledge support from NSF AST-1812642.  SH is grateful for support from the Jack Kent Cooke Foundation.  SH, KC, and DR are grateful to NSF REU grant 1359310 and the Queensborough Community College physics department.   JMB and ML acknowledge support from the CUNY Community College Research Grant. AB acknowledges support from NSF AST-1813871.  CRC acknowledges support from NSF CAREER grant AST-1848107.  FDM and JPS acknowledge support from NSF grant PHY-2013909.   FDM, JPS, and JVN acknowledge support from the University of Oklahoma.  MT is supported by an NSF Astronomy and Astrophysics Postdoctoral Fellowship under award AST-2001810.  JMB thanks Ray Sharma, Nathan Leigh, Amy Reines, Vivienne Baldassare, Jenny Greene, and Mar Mezcua for helpful comments, and Avi Loeb for suggesting the dynamical study described in Section \ref{sect:dynamics}.

Much of our data analysis was done using the {\texttt pynbody}  software suite \citep{pynbody}.  Resources supporting this work were provided by the NASA High-End Computing (HEC) Program through the NASA Advanced Supercomputing (NAS) Division at Ames Research Center.

This research was conducted on Munsee Lenape land.


\bibliographystyle{mnras}
%\bibliography{bh} 

\appendix

\section{Carbon Footprint Calculation}
In order to raise awareness of the substantial impact of High Performance Computing (HPC) to climate change, we calculate the carbon footprint of the eight simulations described in this work (4 DC Justice League and 4 MARVEL-ous Dwarfs; we do not include the isolated galaxies described in Section \ref{sect:dynamics} because this footprint is orders of magnitude smaller.)  We are inspired by the recent work by \citet{Stevens19}, who estimate that among Australian astronomers the carbon footprint of supercomputer use is greater than all other carbon-emitting sources {\em combined}, and is larger than the next-lowest source (travel) by a factor of 4.

The carbon emissions related to a simulation's production depend on several factors, including the power $P_{\rm HPC}$ used by the computer (itself dependent on the number of cores $n_{\rm cores}$ used) and the run time of the simulation $t_{\rm sim}$.  Each of our simulations used slightly different numbers of cores and ran for varying amounts of time; for the sake of simplicity, we do this calculation in an order-of-magnitude fashion.

% https://www.top500.org/system/177259/
The Justice League and MARVEL simulations were run on the Pleiades supercomputer located at NASA Ames Research Center.   The website Top500.org lists the environmental impact of the worlds' leading HPC centers, including information such as power consumption and total number of cores.  Using data from June 2017 (a month when several of our simulations were running),  the NASA HEC Center self-reports a power consumption of 4,407 kW, and hosts 241,108 cores \citep{Top500} for Pleiades.  Assuming equal power consumption across cores, and using the base value of $n_{\rm cores} = 1000$ for each our simulations, we calculate the total power consumption of one simulation to be $P_{\rm HPC} = 18.27$ kW.

Each simulation takes a total of 3-6 months to run on average, but this total includes queue wait times, where no computation occurs.  We use our average value of $t_{\rm sim}$ = 94 days of time for computation.  The total energy expended for one simulation is thus  $P_{\rm HPC} \times t_{\rm sim}$ =  412,354 kWh for one simulation, or 3,298,834 kWh for all eight.

%https://www.epa.gov/energy/greenhouse-gases-equivalencies-calculator-calculations-and-references
We convert energy to amount of CO$_2$ using  the EPA's calculations for home energy use \citep{EPA}, which assumes  an output of 0.454 kg CO$_2$ per kWh of energy.  The total amount of carbon dioxide we have expelled into the Earth's atmosphere with our simulations  equals 1,497,670 kg of CO$_2$.

While there are many ways to offset carbon emissions, a straightforward one is to simply plant trees.   Assuming a typical tree absorbs 158 kg of CO$_2$ in a year \citep{TrueValue}, and lives for 20 years, we calculate that our collaboration should plant 474 trees.  The cost of trees varies greatly by geographic location and species, but is rarely prohibitive.  We encourage researchers to include tree-planting or equivalent carbon offsets in funding proposals.

% Don't change these lines
\bsp	% typesetting comment
\label{lastpage}
\end{document}